\documentclass[twocolumn,showpacs,showkeys,preprintnumbers,amsmath,amssymb]{revtex4} 

\usepackage{graphicx}
\usepackage{dcolumn}
\usepackage{bm}

\begin{document}

\title{Optically monitored nuclear spin dynamics in individual GaAs quantum dots grown by droplet epitaxy}
\author{Thomas\ Belhadj$^1$}
\author{Takashi\ Kuroda$^{2,1}$}
\author{Claire-Marie\ Simon$^1$}
\author{Thierry\ Amand$^1$}
\author{Takaaki\ Mano$^2$}
\author{Kazuaki\ Sakoda$^2$}
\author{Nobuyuki\ Koguchi$^3$}
\author{Xavier\ Marie$^1$}
\author{Bernhard\ Urbaszek$^1$}
\email[Corresponding author : ]{urbaszek@insa-toulouse.fr}

\affiliation{%
$^1$Universit\'e de Toulouse, LPCNO, INSA-CNRS-UPS, 135 avenue de Rangueil, 31077 Toulouse Cedex 4, France}

\affiliation{%
$^2$National Institute for Material Science, Namiki 1-1, Tsukuba 305-0044, Japan}

\affiliation{%
$^3$L-NESS and Dept. Material Science.  Universit\'a de Milano-Bicocca, Via Cozzi 53, I-20125 Milano, Italy}

\date{\today}

\begin{abstract}
We report optical orientation experiments in individual, strain free GaAs quantum dots in AlGaAs grown by droplet epitaxy. Circularly polarized optical excitation yields strong circular polarization of the resulting photoluminescence at 4K. Optical injection of spin polarized electrons into a dot gives rise to dynamical nuclear polarization that considerably changes the exciton Zeeman splitting (Overhauser shift). We show that the created nuclear polarization is bistable and present a direct measurement of the build-up time of the nuclear polarization in a single GaAs dot in the order of one second.   
 
\end{abstract}

\maketitle

\textbf{I. INTRODUCTION}

The spin orientation of carriers injected into nano-meter sized islands of semi-conducting material, quantum dots (QDs), can be probed in spatially resolved optical spectroscopy experiments \cite{Mb2,md2008} or in transport schemes.\cite{hanson07} 
Due to the strong overlap of the electron wave function with a limited number of nuclear spins (10$^4$ to 10$^5$) the hyperfine interaction beween electron and nuclear spins results in stronger effects in QDs than in structures of higher dimensionality. Following first theoretical predictions \cite{Merkulov02,Khaet1} several groups have shown with a large variety of experimental techniques in different dot systems that the electron and nuclear spin system are strongly coupled \cite{pif05,petta05,koppens06,lombez07}. Under most experimental conditions the nuclear spin orientation is to a certain degree random and the effect of the resulting random effective magnetic field leads to electron spin dephasing. Controlling the fluctuations of the nuclear field by achieving a high nuclear polarization (close to 100\%) \cite{Taylor,Christ} or by initializing  the nuclear spins in a dot in a known quantum state is the key to prolonging the electron spin dephasing times in semiconductor QDs, a necessary condition for future coherent and quantum control schemes. \cite{Burkard,Greilich07,hanson07}. Polarization resolved photoluminescence (PL) spectroscopy of single quantum dots allows to monitor simultaneously (i) the electronic polarization by measuring the circular polarization degree of the PL and (ii) the nuclear polarization via changes in the Zeeman splitting (Overhauser shift). The strength and nature of the nuclear effects strongly depend on parameters such as the dot size, the dot material (isotopes), the confinement energy and the strain in the sample.\cite{Aki2006} Detailed studies of strain free GaAs/AlGaAs quantum dots that are formed by interface fluctuations of a GaAs quantum well have revealed strong dynamical nuclear polarization through optical pumping. \cite{GammonPRL,Bracker1} The nuclear polarization in strained InGaAs/GaAs quantum dots formed by self assembly emitting around 950nm  has shown to be bistable. \cite{braun2006,Imapriv,tarta1}

\begin{figure}
\includegraphics[width=0.4\textwidth]{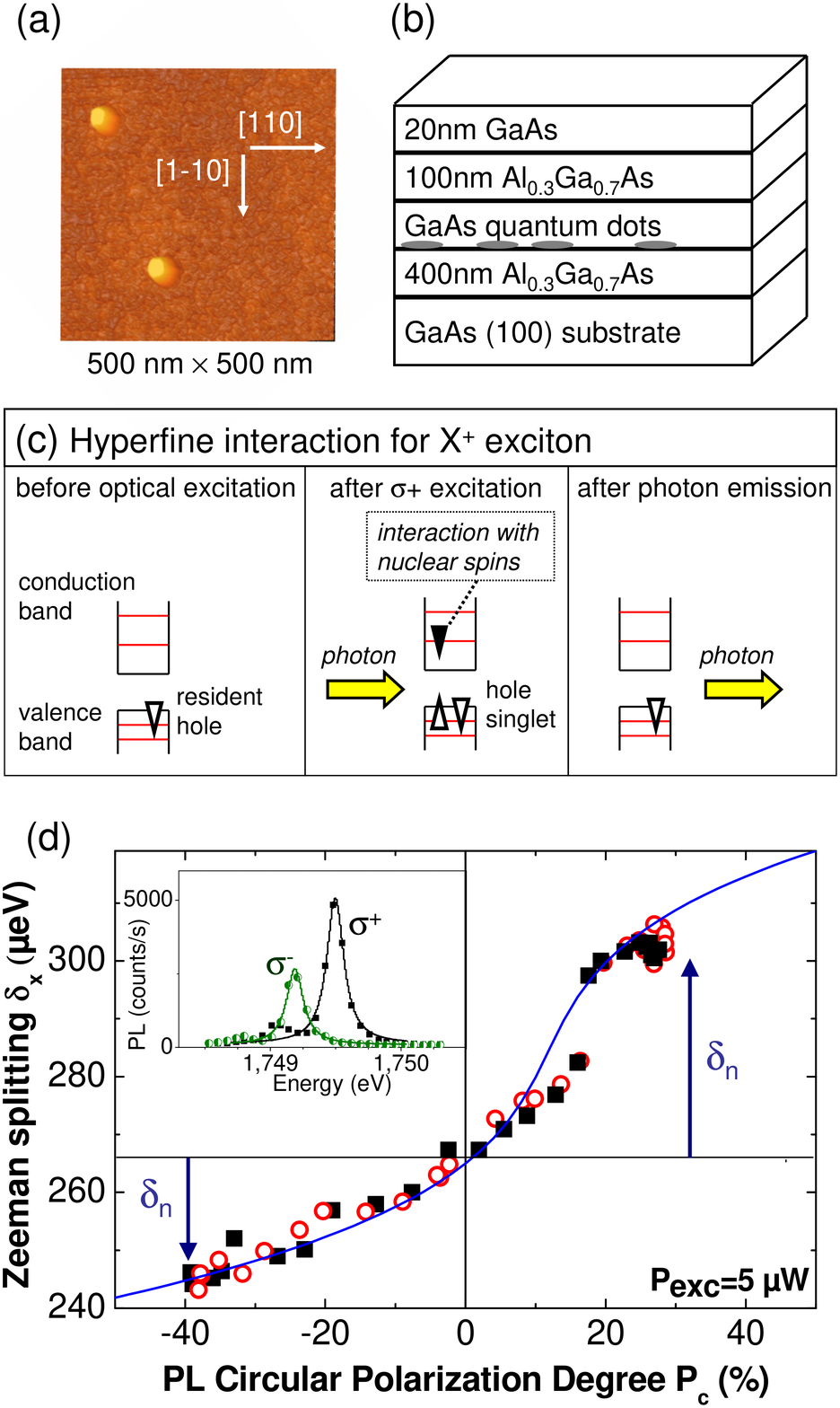}
\caption{\label{fig:fig1} (Color online) (a) Atomic force microscopy image of GaAs dots on the sample surface grown under the same conditions as the burried dots used for optical spectroscopy. (b) Sample structure used for optical measurements. (c) schematic of the X$^+$ exciton, where $\vartriangle$ ($\triangledown$) and $\blacktriangledown$  stand for spin UP (DOWN) heavy holes and spin DOWN electrons, respectively. (d) B$_{ext}$=2.5 T, T=4K. Dot A. The exciton Zeeman splitting is measured as a function of the circular polarization degree of the PL at a constant magnetic field. The solid squares show the measurements when going from low to high electron spin polarization, the hollow circles from high to low. The solid line is a fit with equation (1). Overhauser shift $\delta_n$ indicated by vertical arrow. Inset. A typical PL spectrum is shown for the two different emission polarizations for $\sigma^+$ excitation.}
\end{figure}

Here we report the first optical orientation experiments in a promising system, namely GaAs/AlGaAs dots grown by molecular droplet epitaxy. This system is strain free, contrary to InAs dots, and has a stronger confinement potential than the GaAs interface fluctuation dots, resulting in a total energy difference of 100meV between discrete quantum dot states and delocalized states, compared to typically only 5meV measured for interface fluctuation dots in photoluminescence excitation spectroscopy \cite{hours05}. The physical properties of self assembled quantum dots such as InAs/GaAs and InP/GaInP like the transition energy and the fine structure splitting are determined by the size, the exact material composition and strain effects, all of which vary from dot to dot and can not be measured with 100\% accuracy. All piezoelectric effects due to the strain are absent in GaAs droplet dots. This has several important consequences, and allows, for instance, to study the influence of the quantum dot shape on the fine structure splitting that is crucial for the emission of entangled photon pairs.\cite{abbarchi08} Also, the effects of nuclear depolarization due to quadrupolar coupling are expected to be much weaker than in strained InAs dots.  
We measure in magneto-PL experiments at 4K on single dots the dynamical nuclear polarization created as a function of the optically injected electron spin polarization. A bistability of the nuclear polarization is presented, supported by a direct measurement of the build-up time of the nuclear polarization of 900ms in a single dot.

\textbf{II. EXPERIMENTAL DETAILS}

The strain free GaAs quantum dots can be grown by droplet epitaxy in a molecular beam epitaxy (MBE) machine in AlGaAs barriers in various shapes \cite{Koguchi91,Watanabe00,Kuroda05}. This non-conventional growth method allows quantum dot self assembly in lattice matched systems, i.e. it is not strain driven. The sample investigated here contains the following layers, starting from the substrate: (1) 400nm of $Al_{0.3}Ga_{0.7}As$, (2) GaAs quantum dots, (3) 100nm of $Al_{0.3}Ga_{0.7}As$, (4) 20nm of GaAs, see figure 1b. For atomic force microscopy measurements uncapped dots have been deposited on the sample surface. A typical dot height (diameter) of about 4nm (40nm) is revealed in figure 1a. A low dot density of about 2x10$^8 cm^{-2}$(i.e. 2 per $\mu m^{-2}$) allows single dot measurements without further sample processing.

Single dot PL was carried out with a confocal microscope built around an Attocube nano-positioner connected to a spectrometer and a CCD camera. The narrow linewidth of the PL transitions (limited by the spectral response of our set-up) allowed us to analyze the first optical orientation experiments in single dots grown by droplet expitaxy. Fitting the spectra with a Lorentzian line shape gives a spectral precision of +/- 2.5 $\mu$eV. The spatial resolution of our microscope is given by the detection spot diameter of about 700nm. The circularly polarised cw laser with an energy of 1.95 eV excites carriers non-resonantly in the AlGaAs barrier. The circular polarization degree of the PL is defined as $P_c = (I^+ - I^-)/ (I^+ + I^-)$, where $I^{+(-)}$ is the $\sigma^{+(-)}$ polarized PL intensity. The circular polarization of the laser excitation, defined in a similar way, can be tuned using electrically controlled retardation plates.   

\textbf{III. RESULTS AND DISCUSSION}

The transfer of angular momentum from photons to electron spins and subsequently from electron to nuclear spins has first been reported in 1968 by George Lampel in Silicon \cite{Lampel68}. We first discuss the optical orientation of carrier spins in GaAs droplet dots and show that positively charged excitons X$^+$ (2 valence holes, 1 conduction electron) are created in our sample. The second and third part of the discussion describe in detail the interaction of the electron spin in the X$^+$ exciton with the nuclear spins in the dot.\\ 
\textbf{III.1. Optical orientation of carrier spins in GaAs droplet dots}\\
It is important to note that besides the hyperfine interaction the Coulomb exchange interaction between electron and hole spins determines the exciton polarization Eigenstates in quantum dots and as a consequence the polarization of the emitted photons, for a review see chapter 4 of \cite{md2008}. The anisotropic Coulomb exchange interaction is determining the polarization of a neutral exciton (optically created electron-hole pair) trapped in a quantum dot, masking hyperfine effects at low magnetic fields. The detection of P$_c$ in the order of 20\% at B$_{ext} = 0$ and the strong hyperfine effects discussed in sections III.2 and  III.3 are a clear indication that a majority of the emission lines analyzed in this work originate from the recombination of singly charged excitons (which do emit circularly polarized photons, and not linearly polarized ones). For these three particle complexes the emitted photon polarization following ground state recombinations is \emph{not} determined by the anisotropic exchange interaction, as it cancels out. (see figure 1c). There are two possible cases: the positively charged X$^+$ exciton (1 hole + laser excitation $\Rightarrow$ 2 holes + 1 electron)  and negatively charged exciton X$^-$ (1 electron + laser excitation $\Rightarrow$ 2 electrons + 1 hole) where the resident carriers originate from non-intentional doping.\cite{foot6} Charge tunable structures as for InAs dots \cite{rjw00} do not yet exist for GaAs droplet dots, so we have to identify the charge state by analyzing the optical excitation and the subsequent carrier relaxation process in detail. 

Following optical excitation of the barrier, the initial hole spin orientation is lost due to efficient spin relaxation processes in the barrier, whereas the electron spin orientation is partially conserved during the capture and energy relaxation process in the QD \cite{md2008,Cortez02,Bracker1}. 

The schematic of the X$^+$ exciton ground state is shown in figure 1c.  The two holes are in a singlet state (total spin S=0) and the polarization of the photon emitted by the quantum dot is given by the electron spin $\langle\hat{S}_z^e\rangle=-P_c/2$ \cite{pif05}. During the radiative lifetime of the X$^+$ exciton the electron spin can interact efficiently with the nuclear spins. This interaction will influence the two physical quantities that we measure in our single dot PL experiment: the emitted photon polarization P$_c$ and the photon energy (yielding information about the \emph{nuclear} polarization via changes in the Zeeman energy).  
Following the same argument, we can conclude that we do not deal with the X$^-$ excitons in our sample: In the ground state of the X$^-$, the electron spins form a singlet state (total spin S=0) and the polarization of the emitted photon is determined by the hole spin. The hole spin is orientation is lost in the barrier, so P$_c$ would be averaged to zero in this case, in clear contradiction to our measurements.\cite{foot1} 

\begin{figure}
\includegraphics[width=0.4\textwidth]{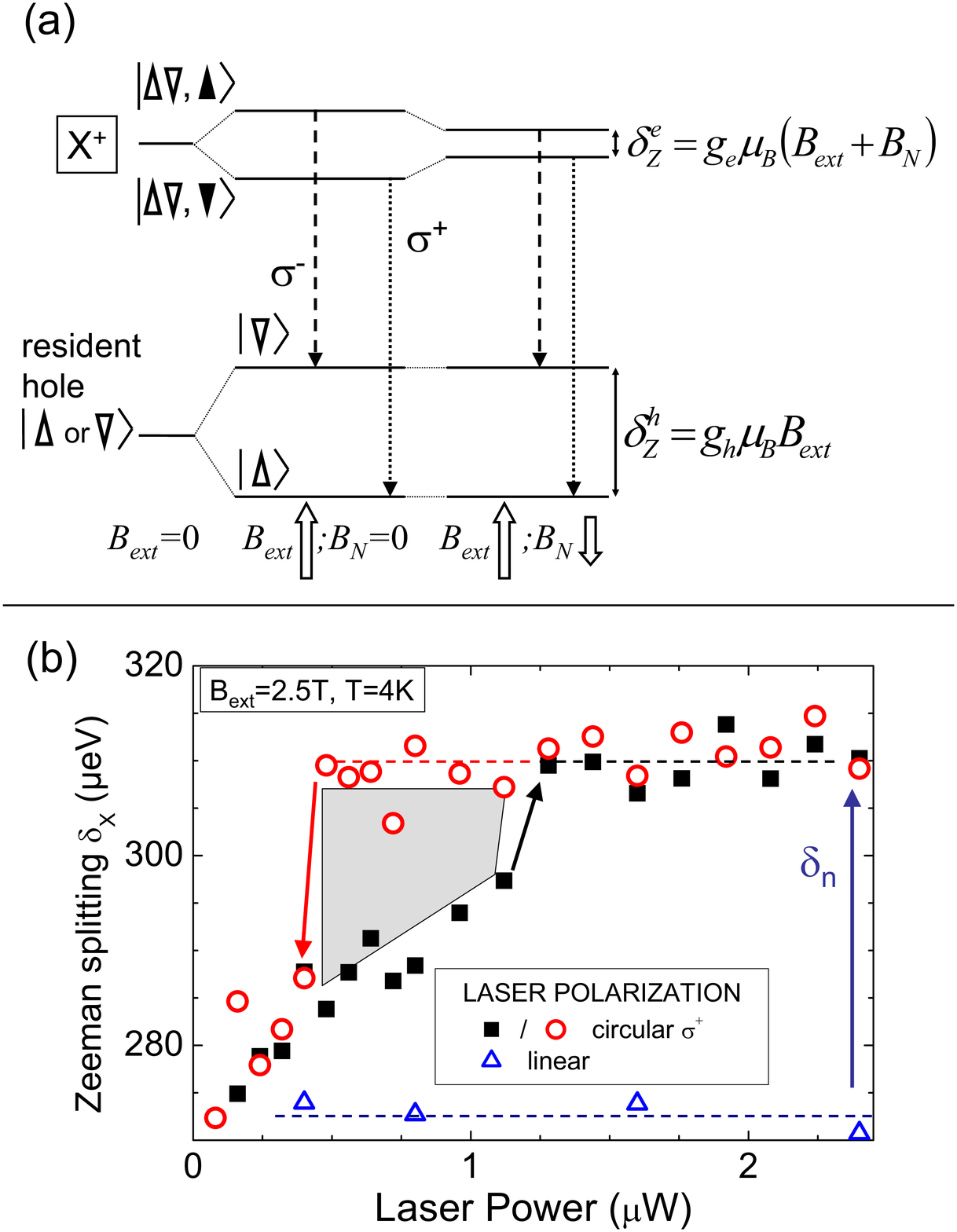}
\caption{\label{fig:fig2} (Color online). (a) Lifting of the degeneracy for both the initial and final state due to the application of B$_{ext}$ and, in the case of electrons, the construction of the nuclear field B$_N$ due to optical pumping. The X$^+$ Zeeman splitting measured in PL is  $\delta_x=\delta_z^h+\delta_z^e$ (b) Dot A. Zeeman splitting $\delta_x$ due to applied field B$_{ext}$ and  effective nuclear field B$_N$. Triangles: $\delta_x$ for linearly polarized excitation stays constant as B$_N \simeq0 $. Squares and circles: $\delta_x$ for $\sigma^+$ excitation. Starting at low pump power, the laser power is first increased (solid squares) and then decreased (hollow circles). Overhauser shift $\delta_n$ indicated by vertical arrows. The signal integration time is several tens of seconds in the (shaded) bistability region ($ P < 1 \mu W$). }
\end{figure}

\textbf{III.2. Optical pumping of nuclear spins and bistability effects}\\ 
In the following we focus on the dynamical nuclear polarization in GaAs QDs in an external magnetic field $B_{ext}$ parallel to the sample growth direction, that is larger than both the local magnetic field $B_L$ (characterising the strength of the nuclear dipole-dipole interaction, in the order of 0.1mT) and the Knight field $B_e$ (the effective magnetic field seen by the nuclei due to the presence of a spin polarized electron, in the order of 10mT \cite{Paget77,Lai,Dzhioev2007}). 
In figure 1d the variation of the exciton Zeeman splitting $\delta_x$ is plotted versus the measured P$_c$ of the PL. The difference $\delta_x (\sigma^{\pm} excitation) - \delta_x (P_c=0)$ is the Overhauser shift $\delta_n$ due to the effective magnetic field B$_N$ created through optical pumping of the nuclear spins as marked by arrows in figure 1d.\cite{foot5}
 A non-linear dependence is observed while only one experimental parameter, the excitation laser polarization, is varied in small steps from $\sigma^-$ to $\sigma^+$. At the origin of this non-linearity, that can even lead to bistability effects, lies the efficient coupling between electron and nuclear spins that we aim to qualitatively describe in what follows by a simple model.
 
The hyperfine interaction between an electron confined to a QD and $N$ nuclei is described by the Fermi contact Hamiltonian. Due to the p-symmetry of the periodic part of the Bloch function the interaction of the hole spin with the nuclear spins is at least one order of magnitude weaker than the interaction with the electron spins. \cite{Abra,Eble2008}. The contribution of the residual doping hole spin to the dynamic nuclear polarization is therefore neglected in the following.  The model developped in reference \cite{Eble} for InAs dots gives an implicit expression for $\delta_n$ and therefore the nuclear polarization as a function of the correlation time of the hyperfine interaction $\tau_c$:

\small
\begin{equation} 
\label{eq:Model}
\delta_n=2\tilde{A}\langle I_z \rangle= -\frac{2\tilde{A}\tilde{Q}\langle \hat{S}_z^e \rangle}{1+\frac{T_{e}(\delta_n)}{T_d}} 
\end{equation}
\normalsize 

where we have introduced $\tilde{A}$ as the average of the hyperfine constants $A^j$ and assuming a strongly simplified, uniform electron wavefunction $\psi(\bar{r})=\sqrt{2/N\nu_0}$ over the involved nuclei and where $\tilde{Q}=\sum_{j}\frac{I^{j}(I^{j}+1)}{S(S+1)}$ and j=As,Ga with a nuclear spin $\hat{I}^j$. We have assumed for simplicity that $T_d$ is an average nuclear decay constant, independent of the nuclear species. The build-up time $T_e$ of the nuclear polarization is: 

\small
\begin{equation}
\label{eq:Te}
T_{e}=\left(\frac{N\hbar}{\tilde{A}}\right)^2\frac{(\frac{\delta_z+\delta_n}{\hbar}\tau_c)^2+1}{2f_e\tau_c}
\end{equation}
\normalsize 

The fraction of time the QD contains an electron is $f_e$ and the electron Zeeman splitting due to B$_{ext}$ is $\delta_z$, where $\delta_z^e=\delta_z+\delta_n$. 

To have a first estimation of the key parameters that determine the nuclear polarization and therefore $\delta_n$ in our dot system, the experimental data in figure 1 is fitted with equation (1) using an electron spin of $\langle\hat{S}_z^e\rangle=-P_c/2$ appropriate for the positively charged exciton X$^+$. The aim of this model is to illustrate the origin of the observed non-linearity: we note that equation (1) has only one solution when $\delta_z = g_e\mu_B B_{ext}$ and $\delta_n = g_e\mu_B B_{N}$ have the same sign, but may have up to three solutions when the signs are opposite. This allows in principle the existence of bistability effects i.e. two stable values of $\delta_n$ for identical experimental conditions, which explains the jump in nuclear polarization in figure 1d at $P_c\approx 20\%$ from one branch to another. For a thorough discussion of nuclear spin bistability and hysteresis effects the reader is referred to chapter 11 of reference \cite{md2008}. \\ 
We can fit the experimental data of figure 1d with equation (1) by varying only $\langle\hat{S}_z^e\rangle$ (the only physical quantity varied in the experiment) with an otherwise fixed set of parameters. Taking into account the large number of parameters, our fit does only give order of magnitude estimates of the physical quantities associated with these parameters.  Three of theses parameter ($T_e,N,g_e$) can be estimated from other experiments: An approximate value of $T_e\simeq 450ms$ is taken, consistent with the measurements in figure 3, see below. $N=10^5$ is chosen according to the approximate size of our dots from the AFM measurements in figure 1a. The electron g-factor in GaAs/AlGaAs quantum wells has been measured as a function of the well thickness in reference \cite{lejeune}. Our dots are 4nm high, so  $g_e=0.2$ is taken as measured for a 4nm thick GaAs quantum well.  This leaves three true fitting parameters: the values of $T_d=2s$, $\tau_c=100ps$ and $f_e=0.05$ were obtained from a least square fit. $T_d$ and $\tau_c$ fit two distinct characteristics of the $\delta_n$ cycle. Adjusting $\tau_c$ allows to fit the width of the bistability region, and in the absence of a bistability, the curvature close to the inflection point of the cycle. $T_d$ determines the maximum nuclear polarization that can be created and fits therefore the extremes of the $\delta_n$ cycle. \cite{bu2007}\\
The nuclear spins in the sample are polarized by flipping their spin simultaneously with the electron spin (flip-flop process) \cite{GammonPRL}. As discussed in detail in references \cite{braun2006,Imapriv,bu2007}, the flip-flop term of the hyperfine interaction is characterized by the correlation time $\tau_c$ and is \emph{switched on} in our experiment only during the existence of the X$^+$ exciton. The uncertainty in the electron Zeeman energy is caracterized by two times: the time it takes the electron to relax towards the quantum dot ground state which we will call capture time $\tau_{cap}$ and the radiative liftetime $\tau_{rad}$. The correlation time $\tau_c$ is dominated by the shorter of these two times when approximating $\tau_c$ by $1/\tau_c=1/\tau_{cap}+1/\tau_{rad}$ . We measured $\tau_{rad}\approx 400ps$ in this sample in time resolved PL experiments, so the 100ps obtained for $\tau_c$ in our fit in figure 1d indicate that $\tau_c$ is determined by  $\tau_{cap}$.\cite{bu2007} 
The relatively short value of $T_d$ could be related to carrier mediated nuclear spin flip mechanisms, that have been discussed for InAs dots \cite{Malet2007}, gate-defined GaAs dots \cite{reilly08} and electrons localized near donors \cite{thierry2008}.

In the case of $\sigma^+$ ($\sigma^-$) excitation the constructed effective nuclear field B$_N$ has to be subtracted from (added to) B$_{ext}$, for dots with a positive electron g-factor (the opposite applies for a negative electron g-factor). As a consequence, flip-flop events are more and more likely under $\sigma^+$ excitation of our sample as the electron Zeeman splitting $\delta_z^e$  decreases further and further, reducing the energy mismatch for a spin flip-flop event (see figure 2a), whereas in the case of $\sigma^-$ excitation there is a negative feedback as $\delta_z^e$ gets bigger. The different electron Zeeman splitting $\delta_z^e$ for the $\sigma^+$ and $\sigma^-$  excitation has led to very different values of B$_N$ in InAs dots, where g$_e\simeq-0.5$ \cite{bu2007}. For the droplet dots we find, similar to GaAs interface fluctuation dots \cite{GammonPRL}, comparable B$_N$ for the two excitation polarizations due to a smaller $|g_e|$ (i.e. the asymmetry is less pronounced than in the case of InAs dots).

We show in figure 2a the Zeeman splitting of the exciton line $\delta_x$ as a function of laser excitation power. Under linearly polarized optical excitation the Zeeman splitting does not change as the average electron spin is close to zero. Under $\sigma^+$ excitation P$_c\simeq30\%$ and the Zeeman splitting first increases significantly with laser power as the nuclear spins are polarized, before a stable value is reached ($P>3 \mu W$), see squares in figure 2b. During the same experimental run we have then decreased the laser power (not changing any other parameter) and the measured $\delta_x$ given by the circles does not follow the same dependence as the squares. Which of the two possible nuclear spin polarizations is reached in the shaded bistability region, does depend on the history of the experiment (non-Markovian behaviour).  Qualitatively similar behaviour has been reported for InAs, InAlAs, InP dots \cite{tarta1,bu2007,kagi07,tarta2}, GaAs quantum wells \cite{Kal92,San2004} and summarized in \cite{md2008}.

\begin{figure}
\includegraphics[width=0.42\textwidth]{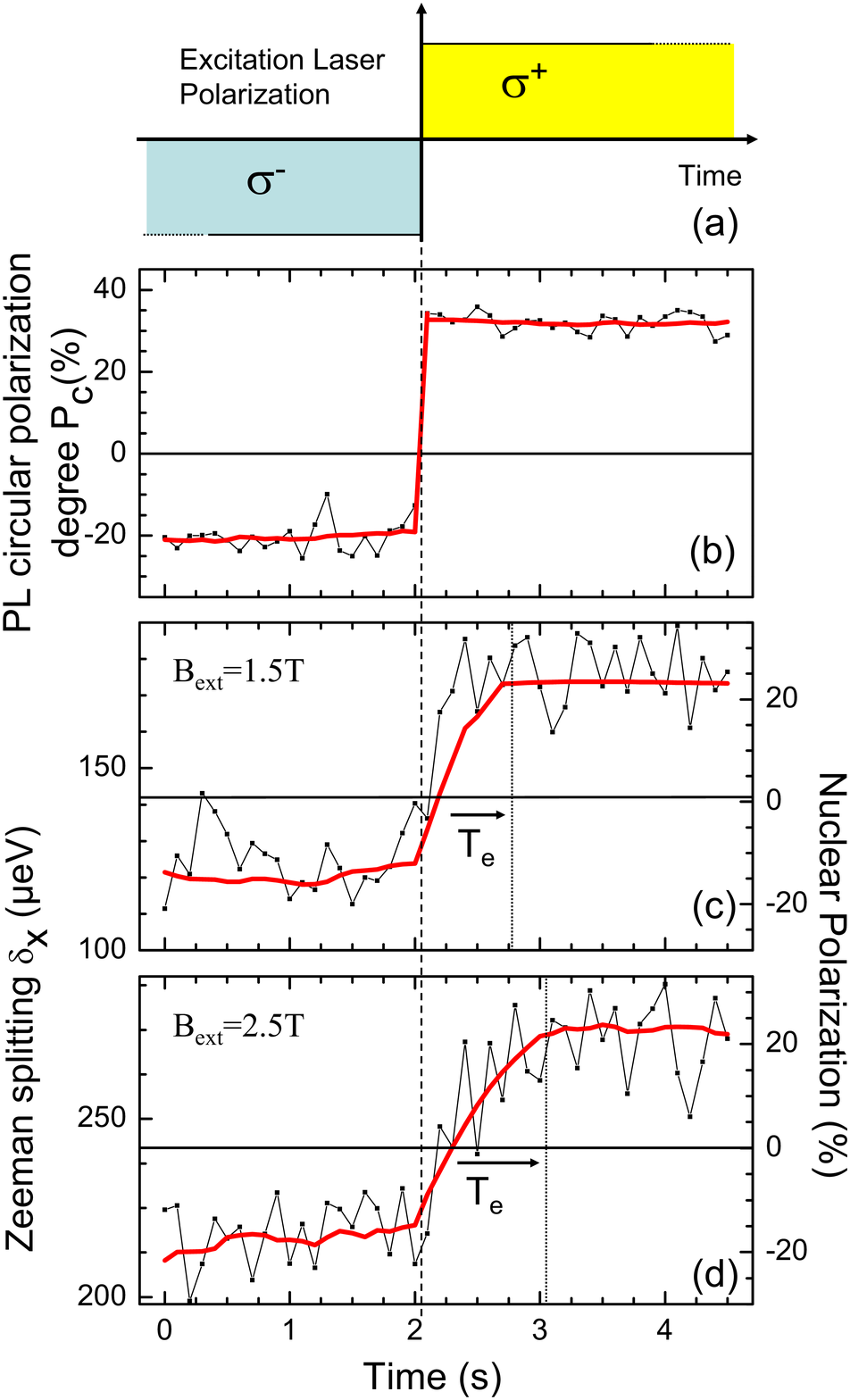}
\caption{\label{fig:fig3} (Color online) , T=4K, P=5$\mu W$. Dot B. (a) The excitation laser polarization is switched electronically from $\sigma^{-}$ to $\sigma^{+}$ at $t_{switch}$ (represented by a dashed vertical line through all panels).  Single dot PL spectra are taken continuously with an acquisition time of about 50ms. (b)B$_{ext}$=2.5 T. The circular polarization degree P$_c$ emitted by a single dot switches orientation at $t_{switch}$ from one acquisition to the next (within 50ms). (c) The Zeeman at splitting at B$_{ext}$=1.5 T (left axis) and hence the nuclear polarization stabilize within 750ms after $t_{switch}$.  Black lines represent the raw data, red lines are averaged over several measurements. (d) same as (c), but forB$_{ext}$=2.5 T}
\end{figure}

\textbf{III.3. Measurement of the build-up time of the nuclear polarization}\\
The results presented in figures 1d and 2b demonstrate the efficient transfer of the spin polarization of an electron trapped in a single GaAs droplet dot to the nuclear spin system. In the following we aim to measure the time needed for the build-up of the nuclear polarization in time resolved experiments. The time resolution of our experiments ($\simeq 50$ms) is essentially the shortest signal integration time that allows the acquisition of a single dot PL spectrum with a tolerable signal to noise ratio and strongly depends on the sample quality, the sensitivity of the CCD detector, etc. In order to investigate the time evolution of the nuclear spin polarization in a single dot we have developed the following measurement protocol : The polarization of the excitation laser is controlled by Meadowlark liquid crystal waveplates, and can be changed computer controlled within about 10ms. For the chosen optical excitation power of 5$\mu W$, the nuclear polarization is maximised (see figure 2b for a typical power dependence). Single dot PL spectra with a signal integration time of about 50ms are recorded continuously and at a certain time $t_{switch}$ after launching the aquisition, the laser polarization changes from $\sigma^-$ to $\sigma^+$, as indicated in figure 3a. After $t_{switch}$, electrons of opposite spin are injected into the dots as confirmed experimentally by the change of sign of $P_c$ from one PL spectrum to the next (figure 3b). In contrast to the electron polarization, the nuclear polarization (figure 3c,d) does not switch instantaneously, but we record a build-up time of the nuclear field in the individual dot in the order of 900ms at B$_{ext}$=2.5T (700ms at 1.5T). The dot in figure 3 is typical for our sample, where T$_e$ is several hundreds of ms for B$_{ext}=1.5 ... 2.5T$, the field range with the largest values of $\delta_n$. To compare this build-up time with various measurements in the literature, one has to keep in mind that in our experiment we do not start from a depolarized nuclear spin ensemble, but in going from $\sigma^-$ to $\sigma^+$ excitation we change orientation of the nuclear field from parallel to anti-parallel with respect to B$_{ext}$. \cite{foot2} We are passing the situation where B$_N$ is zero after about 300ms at B$_{ext}$ when the Zeeman splitting $\delta_x$ measured in figure 3d passes the value $\delta_x(P_c=0)$. 

The build-up time of the nuclear polarization T$_e$ depends directly on the electron Zeeman splitting (given by B$_N$ and B$_{ext}$) and the number of nuclei in the dot, see equation \ref{eq:Te}. Times reported in the literature vary from milliseconds to seconds at B$_{ext}$=0 to 1T in InGaAs dots \cite{Malet2007,Verbin08,Imapriv}, seconds in GaAs interface fluctuation dots at B$_{ext}$=1T, \cite{GammonPRL} to minutes for large gate defined GaAs quantum dots \cite{Tarucha,Foletti}. These much longer times can be qualitatively explained by the lower nuclear spin polarization rate (low value of $f_e$) among other effects like nuclear spin diffusion. In reference \cite{Foletti} the electrons are injected every few micro seconds, compared to every few nano seconds in optical experiments. A clear increase of dynamic nuclear polarization as a function of the electron charging frequency has been found in gate defined GaAs dots \cite{reilly08}. The larger value of N for the dots in these experiments will also increase the build-up time T$_e$.

Measurements in GaAs bulk \cite{Paget82} and in single quantum wells \cite{Harley1,Harley2} have shown that there are essentially two ways of polarizing the nuclear spins. (i) directly via the Fermi contact Hamiltonian. This is very efficient for localized electrons and leads to build-up times in the order of seconds for 50\% of the achievable nuclear polarization detected in a single GaAs quantum well \cite{Harley2}. (ii) Nuclear spins in bulk or quantum wells that are not directly in contact with polarized electrons are polarized via spin diffusion between identical isotopes. In single quantum wells this has lead to a second, much slower build-up time for the remaining nuclei on times scales of 75 to 900 seconds.
As in our experiments the barrier layer is optically excited, it could be speculated that the nuclei in the barrier layer are spin polarized as well as the nuclei of the atoms that form the dot. In this case the single dot emission would just be a nano-metric  probe of a macroscopic nuclear polarization. The fact that we observe a build-up of the steady state value of the nuclear polarization within a few hundred milliseconds is a strong indication that we essentially probe the nuclear polarization created within the dot. \cite{foot3} The barrier states are extended states and as a consequence the rate of nuclear polarization is much slower in the barrier than within the dot. Therefore the build-up time in the barrier is much longer than in the dot (if the existing nuclear depolarization mechanisms allow any build-up of nuclear polarization in the barrier at all). 

Using the polarized nuclei in the GaAs quantum dots as a source for polarizing the nuclear spins in the surrounding AlGaAs matrix via spin diffusion might not be very effective, as the spin diffusion measurement of Malinowski et al \cite{Harley2} between two GaAs quantum wells separated by AlGaAs barriers show. Also in our case the spin diffusion rate from the GaAs dot into the AlGaAs barriers via spin flips of like isotopes which do not involve energy interchange with the lattice will be reduced due to additional quadrupolar splittings, for example due to Al inclusion in the barrier matrix.   
To decide under which conditions (power and duration of optical excitation, sample strain and composition) the decay time of the nuclear polarization in a single dot is influenced by nuclear spin diffusion is a challenge for future experiments \cite{Mak2008}. This is due to the expected long time scales combined with low signal levels and possibly different decay times for different isotopes, as in InGaAs bulk (varying from 6 to 68 minutes). \cite{Kowalski96}

\textbf{IV. CONCLUSION}

To summarize, single dot photoluminescence experiments in GaAs quantum dots in AlGaAs grown by droplet epitaxy show a transfer of a strong, optically generated electron polarization to the nuclear spins in the dot. The time of the initial transfer is measured to be in the range of one second in an external magnetic field of 2.5 T. The strong carrier confinement in these nominally pure GaAs and strain free quantum dots make this an interesting system for studying a single electron spin strongly coupled to nuclear spins. Compared to InAs dots in GaAs the quadropular effects in the samples investigated here are expected to be weak as they will only arise from the small part of the carrier wave function that penetrates the surrounding AlGaAs layer. 

\textbf{V. ACKNOWLEDGEMENTS}

We thank Olivier Krebs and Daniel Paget for fruitful discussions, Pierre-Francois Braun for assistance with the microscope set-up and ANR MOMES and the IUF for financial support.


\begin{thebibliography}{55}
\bibitem[{foo({\natexlab{a}})}]{foot6}
\bibinfo{note}{We can exclude here higher charged states which have far more
  complex emission spectra.}
\bibitem[{foo({\natexlab{b}})}]{foot1}
\bibinfo{note}{It can not be completely excluded at this stage that some
  emission lines with a measurable P$_c$ originate from neutral excitons with
  an anisotropic exchange splitting energy $\delta_1$ of only a few $\mu eV$. A
  measurable $P_c$ can be obtained for neutral excitons if the beat period
  $h/\delta_1$ between angular momentum projections $J_z=+1$ and $J_z=-1$ is
  longer than the radiative lifetime $\tau_{rad}$. We have measured $\tau_{rad}
  \simeq 400 ps$ in this sample, so for a splitting smaller than $h/\tau_{rad}
  \simeq 10 \mu eV$ already a measurable P$_c$ is expected. A detailed
  investigation of the electronic states will be published elsewhere.}
\bibitem[{foo({\natexlab{c}})}]{foot5}
\bibinfo{note}{For many dots in our sample linearly polarized excitation yields
  a positive P$_c$ of a few percent due to thermal polarization at external
  fields of a few Tesla.}
\bibitem[{foo({\natexlab{d}})}]{foot2}
\bibinfo{note}{Due to the low signal to noise ratio for the short signal
  integration time no useful data could be extracted when switching between
  linearly polarized and circularly polarized excitation. We were therefore
  unable to measure the relaxation time of the nuclear spins after going from
  circularly polarized to linearly polarized excitation.}
\bibitem[{foo({\natexlab{e}})}]{foot3}
\bibinfo{note}{Our current signal to noise ratio does not allow us to
  completely exclude the existence of a much slower second component, that adds
  a few percents on a much longer time scale.}
\end{thebibliography}

\end{document}